\documentclass[a4paper,11pt]{article}

\usepackage{graphicx}
\begin{document}

\begin{center}
{\large Nuclear matrix elements of $\beta\beta$ decay from $\beta$-decay data}\par
\bigskip
\bigskip
Jouni Suhonen\footnote{I thank M. Kortelainen for assistance in drawing the figures
of this article. This work has been supported by the Academy of Finland
under the Finnish Centre of Excellence Programme 2000-2005 (Project
No. 44875, Nuclear and Condensed Matter Programme at JYFL).}\par
\bigskip
{\it Department of Physics, University of Jyv\"askyl\"a,
P.O.Box 35, FIN-40014, Jyv\"askyl\"a, Finland}
\end{center}

\begin{abstract}
The evaluation of the nuclear matrix elements (NME) of the
two-neutrino double beta ($2\nu\beta\beta$) decay and
neutrinoless double beta ($0\nu\beta\beta$) decay using the proton-neutron 
quasiparticle random-phase approximation (pnQRPA) is addressed.
In particular, the extraction of a proper value of the proton-neutron 
particle-particle interaction parameter, $g_{\rm pp}$, of this theory is 
analyzed in detail. Evidence is shown, that it can be 
misleading to use the experimental half-life of the $2\nu\beta\beta$ decay 
to extract a value for $g_{\rm pp}$. Rather, arguments are given in favour
of using the available data on single beta decay for this purpose.
\vspace{1pc}
\end{abstract}

\bigskip
\bigskip
\noindent
\textbf{PACS}: 21.60.Jz; 23.40.Hc; 27.50.+e; 27.60.+j\par

\bigskip
\bigskip
\noindent
\textbf{Keywords}: Double beta decay, quasiparticle random-phase approximation,
nuclear matrix elements, proton-neutron interaction

\bigskip
\bigskip
\noindent
\textbf{Status}: Phys. Lett. B, in press

\newpage

The recent large-scale neutrino-oscillation experiments,
Super-Kamioka\-nde \cite{SUP01},
SNO \cite{SNO2002}, KamLAND \cite{KAM2003}, CHOOZ \cite{APP99}, have confirmed
the existence of the neutrino mass. These experiments can only probe the
differences of the squares of the masses, not the absolute mass scale of the
neutrino. On the contrary, the neutrinoless double beta ($0\nu\beta\beta$) decay
can probe the absolute mass scale using the effective neutrino mass,
$\langle m_{\nu}\rangle$, extracted from the results of the
underground double-beta-decay experiments. To extract the absolute 
neutrino masses one needs information about the involved nuclear matrix 
elements \cite{REPORT,Fae98}, neutrino mixing \cite{CIV03}, and
the associated CP phases \cite{PAS02}. As a matter of fact, knowing the 
underlying nuclear matrix elements accurately enough, one can extract from the
double-beta experiments information about the CP phases of the neutrino-mixing
matrix \cite{PAS02}. 

One more fundamental piece of information would emerge if the
$0\nu\beta\beta$ decay were detected, namely that the neutrino would be a
Majorana particle, i.e. an object for whom the particle and antiparticle states
coincide. The $0\nu\beta\beta$ decay then immediately implies also 
nonconcervation of the lepton number, changing the lepton number by two units.
Majorana neutrinos are naturally contained in various 
particle-physics theories going beyond the standard model, 
such as grand-unification theories and supersymmetric 
extensions of the standad model. 

Given the above impressive list of important qualitative and quantitative
neutrino properties, potentially probed by the $0\nu\beta\beta$ decay, one can
not stress enough the importance of a reliable calculation of the involved 
nuclear matrix elements (NME). 
Lack of accuracy in the values of these matrix elements
is the source of inaccuracy in the information on the neutrino masses and CP 
phases, extracted from the $0\nu\beta\beta$-decay experiments. In particular,
in view of the planned near-future large-scale underground experiments, 
with detectors in the ton scale, knowledge of the most promising nuclear 
candidates for detection is of paramount importance.

Contrary to the $0\nu\beta\beta$ decay, the two-neutrino double beta
($2\nu\beta\beta$) decay, with two neutrinos and two electrons in the final state,
can proceed as a perturbative process within the standard model. It can
also be used as a test bench for the nuclear models, since the decay proceeds
via only the $1^+$ states of the intermediate double-odd nucleus. Success in
describing this decay mode is a prerequisite for a reliable calculation of
the NME's related to the $0\nu\beta\beta$ decay.
During the last two decades a host of different nuclear models have been 
used to compute values of the matrix elements involved in both types of
double-beta-decay transition\cite{REPORT,SUH98,SUH02}.
The mostly used nuclear models in the evaluation of the NME's of double
beta decay are the nuclear shell model and the proton-neutron 
quasiparticle random-phase approximation (pnQRPA), designed for spherical or nearly
spherical nuclei. 

After the first shell-model attempts,
the problem of the NME's of the $2\nu\beta\beta$ and $0\nu\beta\beta$ decays
was viewed in a fresh new way by the introduction of the pnQRPA with 
an adjustable particle-particle part of the proton-neutron two-body interaction.
Determination of the value of the corresponding strength parameter, $g_{\rm pp}$,
has been a key issue since the mid 80's. As noticed
in the early works \cite{VOG86,CIV87}, the NME of the $2\nu\beta\beta$ decay is
very sensitive to the value of this parameter, leading to the so-called
$g_{\rm pp}$ problem of the pnQRPA. 
On the other hand, the NME of the $0\nu\beta\beta$ decay is much less dependent
on the value of $g_{\rm pp}$, as discussed, e.g., in \cite{REPORT,SUH92}.
Many extensions of the pnQRPA have come to light during the last nine years.
The first of these was the so-called renormalized pnQRPA (pnRQRPA of Ref. 
\cite{TOI95}). Other extensions of the pnQRPA, used in the $\beta\beta$-decay
calculations, are cited e.g. in \cite{REPORT,Fae98,STO01,ROD03}. 
A common feature of all
these extensions is the attempt to introduce the Pauli exclusion principle
into the pnQRPA by improving on the quasiboson commutation relations, adopted
at the pnQRPA level. In these theories different types of correction to the 
boson commutators of the bifermionic operators are introduced, leading to
renormalization factors at the level of the pnQRPA equations of motion.
 
The results of the $\beta\beta$-decay calculations are quite scattered \cite{SUH04}
(see also \cite{REPORT,SUH98} for a detailed discussion of the matrix elements 
up to the year 1998), and recently it has been suggested that this shortcoming 
could be overcome in the framework of the pnQRPA and its renormalized
extensions. In this
scheme it has been suggested \cite{ROD03} that one could use data on the
$2\nu\beta\beta$ decay to extract a more accurate value for the NME 
corresponding to the neutrino-mass mode (i.e., decay mode mediated by the 
mass of the neutrino) of the $0\nu\beta\beta$ decay. The essentials of
this method are summarized as follows: the value of the interaction strength 
parameter $g_{\rm pp}$ of the pnQRPA (or any of its renormalized extensions) 
can be determined by fitting the value of the computed NME to the one
extracted from the experimental half-life of the corresponding $2\nu\beta\beta$
transition. This fitted value of $g_{\rm pp}$ is then used in the computation 
of the $0\nu\beta\beta$ NME. 
This suggestion has recently been made also in \cite{STO01}. In the following, 
the implications and pitfalls of this scheme are analyzed in detail by using the
simple and transparent framework of the plain pnQRPA. The same qualitative features
persist largely also in its renormalized extensions. At the same time, arguments are
given in favour of an other approach, namely fitting $g_{\rm pp}$ by the data on
single beta decay(s). 

To have an idea of the suggested procedure \cite{ROD03}, and its alternative,
advocated in this work, it is instructive to write down an expression
for the $2\nu\beta\beta$-decay half-life, $t_{1/2}^{(2 \nu)}$, for a transition
from the initial ground state, $0^+_{\rm I}$, to the final ground state, 
$0^+_{\rm F}$. This expression reads
\begin{equation}
\left\lbrack t_{1/2}^{(2 \nu)}(0_{\rm I}^+ \rightarrow 0_{\rm F}^+)
\right\rbrack^{-1}=
G^{(2\nu )}\left\vert M_{\rm DGT}^{(2\nu )}\right\vert^2 \ ,
\label{eq:2vbb}
\end{equation}
where $ G^{(2 \nu)}$  is an integral
over the phase space of the leptonic variables \cite{REPORT}.
The nuclear double Gamow--Teller matrix element, $M_{\rm {DGT}}^{(2 \nu)}$, 
corresponding to the $2\nu\beta\beta$ decay, can be written as
\begin{eqnarray} 
M_{ \rm {DGT}}^{(2 \nu)} & = & \sum_{n} \frac{
(0_{\rm F}^{+} \mid \mid \sum_j \sigma(j) t^{-}_j
\mid \mid 1_n^{+})}{( {{1} \over {2}} Q_{\beta \beta}+
E_n -M_{\rm I})/ m_{\rm e} +1} \times \nonumber \\
 & & (1_n^{+} \mid \mid \sum_j \sigma(j) t^{-}_j
\mid \mid 0_{\rm I}^+)  \; ,
\label{eq:mdgt}
\end{eqnarray}
where the transition operators are the usual Gamow-Teller operators for $\beta^-$
transitions, $Q_{\beta \beta}$ is the $2\nu\beta\beta$ $Q$ value, $E_n$ is the
energy of the $n$th intermediate state, $M_{\rm I}$ is the mass energy
of the initial nucleus, and $m_{\rm e}$ is the rest-mass of the
electron.

As an alternative to the proposed \cite{STO01,ROD03} use of the measured 
$2\nu\beta\beta$ decay half-life to determine the value of $g_{\rm pp}$, 
the use of the measured single-beta-decay half-lives is advocated in this work.
The available data on Gamow--Teller transitions of heavy double-odd 
nuclei, involved in double $\beta^-$ and double $\beta^+$/EC transitions,
have been summarized in Table~\ref{tab:table1}. At the moment, it is believed that
the double $\beta^-$ decays are better accessible to experiments than the
double $\beta^+$/EC decays. Nevertheless, it is instructive to show the
available beta-decay data for nuclei involved in the double $\beta^+$/EC 
decays, as well.

\begin{table}[htb]
\caption{Experimental EC- and $\beta^--\textrm{decay}$ $\log ft$ values for heavy
double-odd nuclei involved as intermediate nuclei in double $\beta^-$ and
double $\beta^+$ decays. For completeness, also the $Q$ values of the
$\beta\beta$ decays are given in the second column.}
\label{tab:table1}
\renewcommand{\arraystretch}{1.2} % enlarge line spacing
\begin{center}
\begin{tabular}{lllllll}\hline
$\beta\beta$ mode & $Q$ [MeV] & Init. nucl. & Final nucl. & Mode & $\log ft$ & Ref. \\
\hline
$\beta^-\beta^-$ & 3.03 & $^{100}$Tc & $^{100}$Mo & EC & 4.45 & \cite{GAR93} \\ 
 & & $^{100}$Tc & $^{100}$Ru & $\beta^-$ & 4.6 & \cite{FIR96} \\ 
$\beta^-\beta^-$ & 1.30 & $^{104}$Rh & $^{104}$Ru & EC & 4.3 & \cite{FIR96} \\ 
 & & $^{104}$Rh & $^{104}$Pd & $\beta^-$ & 4.5 & \cite{FIR96} \\ 
$\beta^+\beta^+$ & 0.73 & $^{106}$Ag & $^{106}$Pd & EC & 4.9 & \cite{FIR96} \\ 
 & & $^{106}$Ag & $^{106}$Cd & $\beta^-$ & $\ge 4.2$ & \cite{FIR96} \\ 
$\beta^-\beta^-$ & 2.01 & $^{110}$Ag & $^{110}$Pd & EC & 4.1 & \cite{FIR96} \\ 
 & & $^{110}$Ag & $^{110}$Cd & $\beta^-$ & 4.7 & \cite{FIR96} \\ 
$\beta^-\beta^-$ & 0.53 & $^{114}$In & $^{114}$Cd & EC & 4.9 & \cite{FIR96} \\ 
 & & $^{114}$In & $^{114}$Sn & $\beta^-$ & 4.5 & \cite{FIR96} \\ 
$\beta^-\beta^-$ & 2.80 & $^{116}$In & $^{116}$Cd & EC & 4.39 & \cite{BHA98} \\ 
 & & $^{116}$In & $^{116}$Sn & $\beta^-$ & 4.7 & \cite{FIR96} \\ 
$\beta^-\beta^-$ & 0.87 & $^{128}$I & $^{128}$Te & EC & 5.0 & \cite{FIR96} \\ 
 & & $^{128}$I & $^{128}$Xe & $\beta^-$ & 6.1 & \cite{FIR96} \\ 
$\beta^+\beta^+$ & 0.54 & $^{130}$Cs & $^{130}$Xe & EC & 5.1 & \cite{FIR96} \\ 
 & & $^{130}$Cs & $^{130}$Ba & $\beta^-$ & 5.1 & \cite{FIR96} \\ 
$\beta^+\beta^+$ & 0.37 & $^{136}$La & $^{136}$Ba & EC & 4.6 & \cite{FIR96} \\ 
 & & $^{136}$La & $^{136}$Ce & $\beta^-$ & ? & \cite{FIR96} \\ 
\hline
\end{tabular} 
\end{center}
\end{table} 

As the first, clean-cut test case one can take the 
decay of $^{116}$Cd which is a nearly spherical, almost semi-magic nucleus. 
The corresponding final nucleus of the $2\nu\beta\beta$ decay is $^{116}$Sn, a 
genuine spherical semi-magic nucleus. Both these nuclei are well 
describable by the spherical pnQRPA.

The calculation of the matrix element of Eq.~(\ref{eq:mdgt}) proceeds on the
following lines. The single-particle energies of the sherical mean field
are obtained from a Woods--Saxon single-particle potential, including the
Coulomb and spin-orbit parts in the Bohr--Mottelson parametrization \cite{BOH69}.
The single-particle valence space is taken typically to span two to three 
oscillator major shells around the proton and neutron Fermi surfaces. The
adopted two-body interaction is a realistic one, based on the
one-boson-exchange potential of the Bonn type, transformed to nuclear matter
by the G-matrix technique. The finite-size effects have been taken into account
in an approximate way by using simple scaling parameters for the short-range
monopole part, and separate scalings for the $J^{\pi}= 1^+$ multipole in the
particle-hole and particle-particle channels. Details of the calculation can
be read in \cite{CIV99}.

The strong short-range correlations between nucleons have been treated by 
using the BCS approximation. The associated pairing strengths are adjusted
to reproduce the empirical pairing gaps, extracted from the experimental
separation energies of protons and neutrons, in a way described in \cite{SUH88}.
The proton-neutron correlations are treated at the pnQRPA level by fixing the
scale of the particle-hole $J^{\pi}= 1^+$ two-body matrix elements to reproduce
the empirical location of the Gamow--Teller giant resonance, whereas the
particle-particle part of the same interaction is scaled by the interaction
strength constant $g_{\rm pp}$, left as a free parameter in the calculations.
This method was used for realistic interactions in the context of the
$2\nu\beta\beta$ decay in \cite{CIV87}, and in description of single beta
decays in \cite{SUH88}.

\begin{figure}[htb]
\begin{center}
\includegraphics[width=13cm]{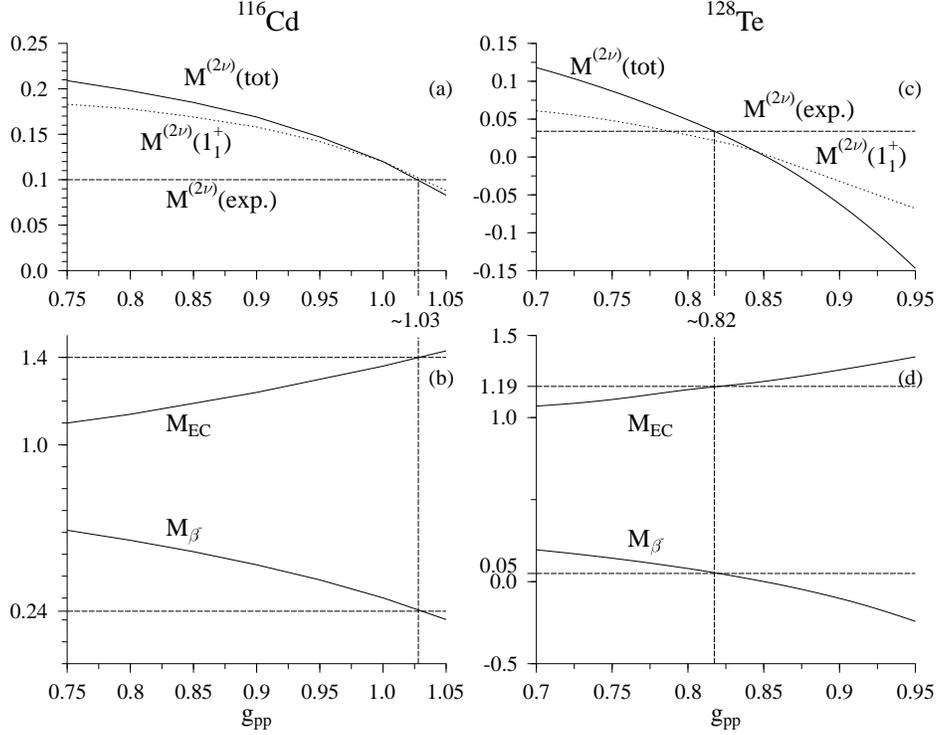}
\end{center}
\caption{Panel (a): The NME's corresponding to the $2\nu\beta\beta$ decay of
$^{116}$Cd shown as functions of $g_{\rm pp}$.
The complete NME, $M^{(2\nu )}(\textrm{tot})$, the NME with only the lowest
intermediate contribution included, $M^{(2\nu )}(1^+_1)$, and the experimental
NME, $M^{(2\nu )}(\textrm{exp.})$, have been shown. Panel (b): The left-branch,
EC NME, $M_{\rm EC}$, and the right-branch NME, $\beta^-$ NME, $M_{\beta^-}$,  
shown as functions of $g_{\rm pp}$. Panels (c) and (d): The same as (a) and (b)
for the $2\nu\beta\beta$ decay of $^{128}$Te.}
\label{fig:gpp-cdte}
\end{figure}

In Fig.~\ref{fig:gpp-cdte}, panel (a), the NME $M^{(2\nu )}(\textrm{tot})$, 
corresponding to the $2\nu\beta\beta$ decay of
$^{116}$Cd, is drawn as a function of $g_{\rm pp}$. In the same figure a 
rough value of the extracted experimental NME, $M^{(2\nu )}(\textrm{exp.})$,
has been shown as a horizontal line, since its
value is independent of $g_{\rm pp}$. Here the uncertainties in the value of this
extracted NME, arising from the experimental error in the measured half-life, and
the uncertainty in the proper value of the axial-vector coupling constant, 
$g_{\rm A}$, for
medium-heavy and heavy nuclei, have been omitted. The intersection point of
these two curves gives now the fitted value, $g_{\rm pp}(\beta\beta)\simeq 1.03$, of
$g_{\rm pp}$. As can be seen from the curve denoted by $M^{(2\nu )}(1^+_1)$ in
Fig.~\ref{fig:gpp-cdte}, the NME including only the contribution 
arising from the virtual 
transition through the first $1^+$ state, $1^+_1$, of the intermediate nucleus 
$^{116}$In, almost coincides with the complete NME, $M^{(2\nu )}(\textrm{tot})$,
especially for $g_{\rm pp}$ values around unity. This is a characteristic of
the so-called single-state dominance (SSD), studied extensively e.g. in 
\cite{CIV99}. 

In the case of such a SSD, the NME (\ref{eq:mdgt}) of the 
$2\nu\beta\beta$ decay can be approximately written as
\begin{equation}
M^{(2\nu )} \simeq \frac{M_{\rm EC}M_{\beta^-}}{( {{1} \over {2}} Q_{\beta \beta}+
E_1 -M_{\rm I})/ m_{\rm e} +1} \ .
\label{eq:2vme}
\end{equation}
The two branches of the $2\nu\beta\beta$ transition, $M_{\rm EC}$ and
$M_{\beta^-}$ are drawn as functions of $g_{\rm pp}$ in panel (b) of 
Fig.~\ref{fig:gpp-cdte}. It is remarkable that the magnitudes of the 
left-branch NME, 
corresponding to the electron-capture (EC) decay of the $1^+_1$ state in 
$^{116}$In to the ground state of $^{116}$Cd, and the right-branch NME, 
corresponding to the $\beta^-$ decay of the same state to the ground state 
of $^{116}$Sn, can in some cases
be determined from experimental data on the corresponding
decay half-lives. It is also clear that in this kind of a simple case the study
of the relation between the single and double beta decays is most transparent,
in particular, related to the determination of the $g_{\rm pp}$ parameter.

Using the extracted value of $g_{\rm pp}(\beta\beta)$, one immediately obtains,
due to the SSD, the values of the left- and right-branch NME's, as shown in
panel (b) of Fig.~\ref{fig:gpp-cdte}. 
From Fig.~\ref{fig:gpp-cdte} one obtains $M_{\rm EC}\simeq 1.4$ and
$M_{\beta^-}\simeq 0.24$. These values of the NME's can, in turn, be used to
compute the half-lives of the EC and $\beta^-$ decays from the $1^+_1$ state in
$^{116}$In. Comparison of these computed values with the corresponding 
experimental ones, extracted from Table~\ref{tab:table1}, yields
\begin{equation}
\frac{t_{1/2}^{(\textrm{EC})}(\textrm{exp.})}{t_{1/2}^{(\textrm{EC})}(\textrm{th.})}
\simeq 2.6\quad ;\quad
\frac{t_{1/2}^{(\beta^-)}(\textrm{exp.})}{t_{1/2}^{(\beta^-)}(\textrm{th.})}
\simeq 0.16 \ ,
\label{eq:half-cd}
\end{equation}
indicating that for $g_{\rm pp}(\beta\beta)\simeq 1.03$ one obtains too fast an
EC transition and much too slow a $\beta^-$ transition. Fitting the $\beta^-$
decay half-life, instead of the $2\nu\beta\beta$ decay half-life, would yield
a value $g_{\rm pp}(\beta^-)\simeq 0.85$, which also would result in a more
reasonable matrix element for the EC branch, namely $M_{\rm EC}\simeq 1.2$.
The corresponding experimental magnitude is $M_{\rm EC}(\textrm{exp})\simeq 0.8$,
the exact value depending on the adopted value for $g_{\rm A}$.
As can be seen, the proper determination of the value of the
$g_{\rm pp}$ parameter, by using the $\beta^-$ decay half-life, can lead to a
notably different value from the one extracted by using the $2\nu\beta\beta$ 
decay half-life, even in the simple case of the SSD. Summarizing the above:
use of the value $g_{\rm pp}(\beta\beta)\simeq 1.03$ reproduces the 
$2\nu\beta\beta$ half-life via two compensating errors: too large an EC NME 
is compensated by too small a $\beta^-$ NME.

As the second test case one can take the $2\nu\beta\beta$ decay of $^{128}$Te 
to the ground state of $^{128}$Xe. This case can be analyzed using the very 
methods deviced for $^{116}$Cd in Fig.~\ref{fig:gpp-cdte}, panels (a) and (b). 
A corresponding scheme is shown for the $^{128}$Te decay
in Fig.~\ref{fig:gpp-cdte}, panels (c) and (d). 
As can be seen from panel (c), the 
curves for the total matrix element and the $M^{(2\nu )}(1^+_1)$ matrix 
element are very much separated everywhere but at the values of $g_{\rm pp}$ 
close to the point which reproduces the value of the experimental matrix 
element. Hence, in this case one can not speak about SSD, and the situation 
is more complicated than in the $^{116}$Cd case. In this case the 
intersection point of the curves, corresponding to the total and experimental
matrix elements, gives the fitted value, $g_{\rm pp}(\beta\beta)\simeq 0.82$, of
$g_{\rm pp}$. 

The two branches of the matrix element $M^{(2\nu )}(1^+_1)$,
$M_{\rm EC}$ and $M_{\beta^-}$, are drawn as functions of $g_{\rm pp}$ in panel 
(d) of Fig.~\ref{fig:gpp-cdte}.
Using the extracted value of $g_{\rm pp}(\beta\beta)$, one obtains for the 
left- and right-branch matrix elements $M_{\rm EC}\simeq 1.19$ and
$M_{\beta^-}\simeq 0.05$. These values of the NME's can, in turn, be used to
compute the half-lives of the EC and $\beta^-$ decays from the $1^+_1$ state of
$^{128}$I. Comparison of these computed values with the corresponding 
experimental ones, extracted from Table~\ref{tab:table1}, yields
\begin{equation}
\frac{t_{1/2}^{(\textrm{EC})}(\textrm{exp.})}{t_{1/2}^{(\textrm{EC})}(\textrm{th.})}
\simeq 9.7\quad ;\quad
\frac{t_{1/2}^{(\beta^-)}(\textrm{exp.})}{t_{1/2}^{(\beta^-)}(\textrm{th.})}
\simeq 0.17 \ ,
\label{eq:half-te}
\end{equation}
indicating that for $g_{\rm pp}(\beta\beta)\simeq 0.82$ one obtains much too fast 
an EC transition and much too slow a $\beta^-$ transition. Fitting the $\beta^-$
decay half-life, instead of the $2\nu\beta\beta$ decay half-life, would yield
a value $g_{\rm pp}(\beta^-)\simeq 0.755$, which would only slightly change the
value of the matrix element for the EC branch, namely to $M_{\rm EC}\simeq 1.15$.
The corresponding experimental magnitude is $M_{\rm EC}(\textrm{exp})\simeq 0.38$,
for $g_{\rm A}=1.0$. As can be seen, in this case the proper determination 
of the value of the $g_{\rm pp}$ parameter, by using the $\beta^-$ decay 
half-life, does not lead to a notably different value of $M_{\rm EC}$
from the one extracted by using the $2\nu\beta\beta$ decay half-life. 
The reason for this discrepancy is not clear, but deformation effects could
play some role. Even so, the above tells us that the
use of the value $g_{\rm pp}(\beta\beta)\simeq 0.82$ reproduces the 
$2\nu\beta\beta$ half-life via two compensating errors: too large an EC NME 
is compensated by too small a $\beta^-$ NME.

\begin{figure}[htb]
\begin{center}
\includegraphics[width=13cm]{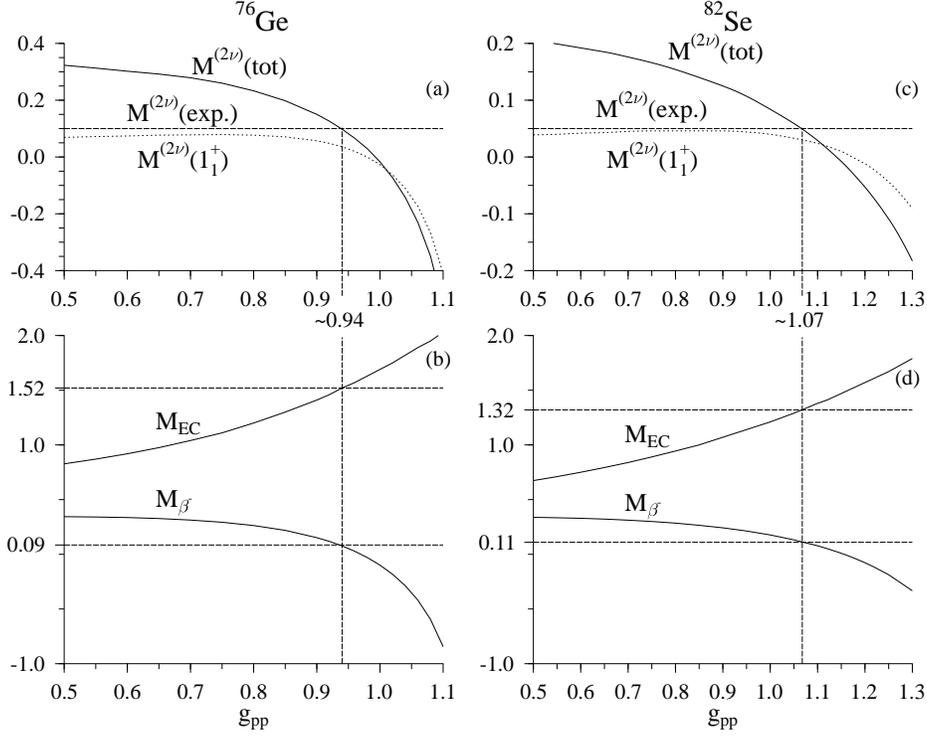}
\end{center}
\caption{The same as Fig.~\ref{fig:gpp-cdte} for NME's corresponding 
to the $2\nu\beta\beta$ decays of $^{76}$Ge and $^{82}$Se.}
\label{fig:gpp-gese}
\end{figure}

As the third case, the $2\nu\beta\beta$ decay of $^{76}$Ge to the ground state of
$^{76}$Se will be discussed. 
This case can be analyzed along the lines of the previous two cases.
A corresponding scheme is shown for the $^{76}$Ge decay in Fig.~\ref{fig:gpp-gese},
panels (a) and (b). As can be seen from Fig.~\ref{fig:gpp-gese}, there exists no
SSD, and the situation is in this respect similar to the $^{128}$Te case. 
In this case the 
intersection point of the curves, corresponding to the total and experimental
matrix elements, gives the fitted value, $g_{\rm pp}(\beta\beta)\simeq 0.94$, of
$g_{\rm pp}$. 

The two branches of the matrix element $M^{(2\nu )}(1^+_1)$,
$M_{\rm EC}$ and $M_{\beta^-}$, are drawn as functions of $g_{\rm pp}$ in panel 
(b) of Fig.~\ref{fig:gpp-gese}. Using the extracted value of 
$g_{\rm pp}(\beta\beta)$, one obtains the values $M_{\rm EC}\simeq 1.52$ and 
$M_{\beta^-}\simeq 0.09$ of the left- and right-branch NME's, which give 
for the corresponding $\log ft$ values
\begin{equation}
\log ft(\textrm{EC})\simeq 3.9\quad ;\quad \log ft(\beta^-)\simeq 6.4 \ ,
\label{eq:half-ge}
\end{equation}
for $g_{\rm A}= 1.0$. The lowest state in the intermediate nucleus, $^{76}$As,
is a $2^-$ state, and hence the Gamow--Teller decays of the lowest $1^+$ state 
are hard to observe due to the fast gamma decays to this $2^-$ state. 

As the next example of the $g_{\rm pp}(\beta\beta)$ problem, 
the $2\nu\beta\beta$ decay of $^{82}$Se to the ground state of $^{82}$Kr is
discussed in Fig.~\ref{fig:gpp-gese}, panels (c) and (d). 
As for the $^{76}$Ge case, also here the SSD is not
applicable. From Fig.~\ref{fig:gpp-gese} one can read for the intersection point
of the total and experimental matrix elements the value 
$g_{\rm pp}(\beta\beta)\simeq 1.07$, giving for the EC and $\beta^-$ NME's the
values $M_{\rm EC}\simeq 1.32$ and $M_{\beta^-}\simeq 0.11$. 
These, in turn, give for the corresponding $\log ft$ values
\begin{equation}
\log ft(\textrm{EC})\simeq 4.0\quad ;\quad \log ft(\beta^-)\simeq 6.2 \ ,
\label{eq:half-se}
\end{equation}
for $g_{\rm A}= 1.0$. The lowest two states in the intermediate nucleus, $^{82}$Br,
are a $5^-$ state and a $2^-$ state, and hence the Gamow--Teller decays of 
the lowest $1^+$ state have not been observed. 

\begin{table}[htb]
\caption{EC- and $\beta^--\textrm{decay}$ $\log ft$ values for selected decays of
double-odd nuclei in the pf shell. The data is taken from \cite{FIR96}.}
\label{tab:table2}
\renewcommand{\arraystretch}{1.2} % enlarge line spacing
\begin{center}
\begin{tabular}{llll}\hline
Init. nucl. & Final nucl. & Mode & $\log ft$ \\
\hline
$^{70}$Ga & $^{70}$Zn & EC & 4.7 \\ 
$^{70}$Ga & $^{70}$Ge & $\beta^-$ & 5.1 \\ 
$^{78}$Br & $^{78}$Se & EC & 4.8 \\ 
$^{78}$Br & $^{78}$Kr & $\beta^-$ & ? \\ 
$^{80}$Br & $^{80}$Se & EC & 4.7 \\ 
$^{80}$Br & $^{80}$Kr & $\beta^-$ & 5.5 \\ 
\hline
\end{tabular} 
\end{center}
\end{table} 

Although no measured EC or $\beta^-$ NME can be extracted for the $^{76}$Ge and
$^{82}$Se cases, one can compare the computed $\log ft$ values of 
Eqs.~(\ref{eq:half-ge}) and (\ref{eq:half-se}) to the $\log ft$ values of 
similar cases in the same
mass region. In the relevant mass region there are three double-odd nuclei with
a $1^+$ ground state and decay patterns analogous to the ones of $^{76}$As 
and $^{82}$Br, namely the ones listed in Table~\ref{tab:table2}. From this table
one immediately notices that the $\log ft$ values of the relevant EC decays
range between $\log ft(\textrm{EC}) = 4.7-4.8$ and values of the relevant
$\beta^-$ decays range between $\log ft(\beta^-) = 5.1-5.5$. This would
suggest that the extracted $\log ft$ values of Eqs.~(\ref{eq:half-ge}) and 
(\ref{eq:half-se}) for the EC decays are too small and the corresponding
extracted $\log ft$ values for the $\beta^-$ decays far too large. Much better
agreement between the theoretical and experimental EC and $\beta^-$ 
$\log ft$ values, around $\log ft(\textrm{EC}) \simeq 4.6$ and 
$\log ft(\beta^-) \simeq 5.3$ could be obtained for smaller $g_{\rm pp}$ values
than the one, suggested by the $2\nu\beta\beta$-decay half-life. 
For the two discussed
decays a value $g_{\rm pp}(\beta^-) \simeq 0.8$ would do quite well.

Based on the previous analysis one can say that the conclusions arising from 
the analysis of the $2\nu\beta\beta$ decays of $^{76}$Ge and $^{82}$Se 
coincide with the ones
arising from the $2\nu\beta\beta$ decays of $^{116}$Cd and $^{128}$Te: 
cancellation of errors in the two matrix elements, $M_{\rm EC}$ and 
$M_{\beta^-}$, conspire to produce a $2\nu\beta\beta$ NME which exactly 
reproduces the corresponding experimental matrix element. 
This demonstrates that it can be dangerous to determine the value of 
$g_{\rm pp}$ by fitting the $2\nu\beta\beta$ decay half-life. 

In fact, determination of the value of $g_{\rm pp}$ by the data on
single beta decay leaves the $2\nu\beta\beta$-decay half-life as a prediction
of the theory. Comparison of this prediction to the experimental half-life
would tell about the predictive power of the adopted theoretical framework,
in terms of the size of the adopted single-particle space, the adopted
single-particle energies, etc. A rougly correct prediction for the 
$2\nu\beta\beta$-decay half-life would shed more confidence on the theoretical
predictions concerning the other multipoles, involved in the $0\nu\beta\beta$
decay. 

\begin{figure}[htb]
\begin{center}
\includegraphics[width=8cm]{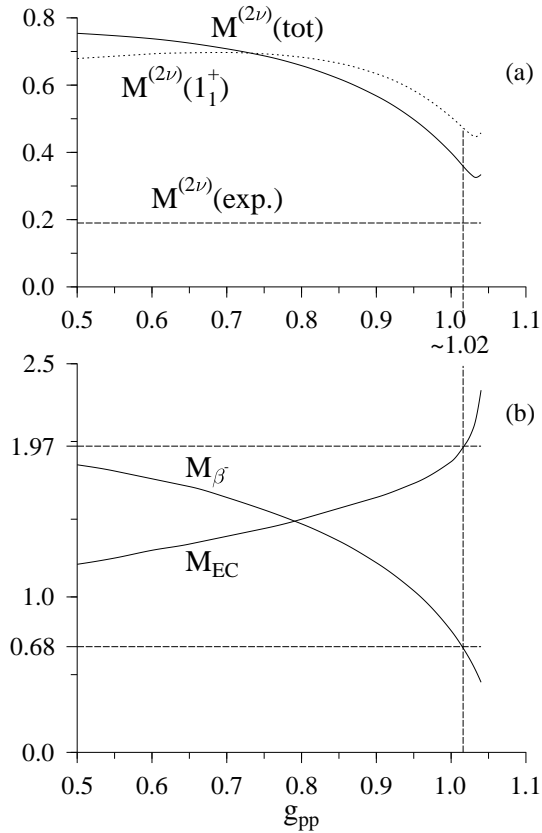}
\end{center}
\caption{The same as Fig.~\ref{fig:gpp-cdte} for NME's corresponding 
to the $2\nu\beta\beta$ decay of $^{100}$Mo.}
\label{fig:gpp-mo}
\end{figure}

As the final example of the $g_{\rm pp}(\beta\beta)$ problem, 
the $2\nu\beta\beta$ decay of $^{100}$Mo to the ground state of $^{100}$Ru is
discussed in Fig.~\ref{fig:gpp-mo}. In this case the SSD is roughly applicable.
From Fig.~\ref{fig:gpp-mo} one can see that in this particular nuclear-structure
calculation, the one of Ref. \cite{CIV99} where one can read more details of
the used single-particle basis, etc., the computed total NME never reaches the
experimental NME, extracted by using $g_{\rm A}= 1.0$. Hence, in this case one
is forced to use the experimental $\beta^-$-decay $\log ft$ value, quoted in
Table~\ref{tab:table1}, to determine the value of $g_{\rm pp}$, resulting in
$g_{\rm pp}(\beta^-)\simeq 1.02$. This gives for the EC the value 
$M_{\rm EC}\simeq 1.97$, and for the corresponding $\log ft$ value
\begin{equation}
\log ft(\textrm{EC})\simeq 3.7 \ ,
\label{eq:half-mo}
\end{equation}
using $g_{\rm A}= 1.0$. This is too low a value for this decay, as seen from
the data of Table~\ref{tab:table1}, indicating that some nuclear-structure
effects, e.g. deformation, beyond the reach of the spherical pnQRPA, might 
be present.

\begin{table}[ht!]
\caption{Pros (+) and cons ($-$) of the two discussed recipes to fit the parameter 
$g_{\rm pp}$. For more explanation on the various points see the text.}
\label{tab:table3}
\renewcommand{\arraystretch}{1.2} % enlarge line spacing
\begin{center}
\begin{tabular}{lll}\hline
Point & Fit to $\beta^-$ and/or EC decay(s) & Fit to $2\nu\beta^-\beta^-$ decay \\
\hline
1 & One, two or more observables & Only one observable can be \\
 & can be used for the fit (+) & used for the fit ($-$) \\
\hline
2 & Direct access to grass-root-level & Two or more compensating \\
 & deficiencies of a nuclear model (+) & errors may conspire to produce \\
 & & a good $2\nu\beta^-\beta^-$ decay rate ($-$) \\
\hline
3 & The beta-decay properties better & The $2\nu\beta^-\beta^-$ decay
properties \\
 & reproduced (+) & better reproduced (+) \\
\hline
4 & Error limits from comparison  & Advisable to check against \\
 & of the experimental and computed & data on $\beta^-$ decays \\
 & $2\nu\beta^-\beta^-$ decay rate (+) &  \\
\hline
5 & Largely eliminates the model-space & Largely eliminates the 
model-space \\
 & dependence of the computed & dependence of the computed \\
 & $0\nu\beta^-\beta^-$ decay rates (+) & $0\nu\beta^-\beta^-$ decay rates (+) \\
\hline
6 & Can be extended to study of & No access to a possible variation \\
 & forbidden contributions, & of $g_{\rm pp}$
from multipole \\
 & e.g. $2^-$, in $0\nu\beta^-\beta^-$ decay (+) & to multipole ($-$) \\
\hline
7 & Can access $\beta\beta$ decays where no & Can access 
$\beta\beta$ decays where no \\
 & $2\nu\beta\beta$ data exists, see & direct $\beta$-decay data exists \\
 & Table~\ref{tab:table2} (+)  & ($^{76}$Ge and $^{82}$Se) (+) \\
\hline
Balance & $7\times (+)$ & $3\times (+)$ and $3\times (-)$ \\
\hline
\end{tabular} 
\end{center}
\end{table} 

Summarizing the above presented results, the problem of
determination of the proton-neutron interaction strength, $g_{\rm pp}$,
in a pnQRPA type of calculation, be it the plain pnQRPA or one of its 
renormalized extensions, has been addressed. The apparent solution
of the ``$g_{\rm pp}$ problem'' by fitting $g_{\rm pp}$ to available
data on $2\nu\beta\beta$-decay half-lives has been critically analyzed. Fitting
this parameter to the existing data on single $\beta^-$ transitions is found
to be a more meaningful solution to the problem. Arguments favouring this method
have been summarized in Table~\ref{tab:table3} where the positive points (+) and
negative points ($-$) of the two fitting methods have been listed. Below few comments
concerning the listed points of the table are made.

Concerning point one, one can even perform a systematical study
of the beta-decay properties of a given nuclear region in the fit to
beta decays. This approach would correspond to a shell-model \cite{CAU96} 
type of application of the beta-decay data. Referring to point two, a fit to $\beta^-$
data can reveal deficiencies in the predictive power of the used nuclear model 
in the case of the EC rates of the other branch. This seems to be the 
case, e.g., in the present calculation. 

In regard to point four, the first method can be used to draw some
conclusions about the error limits in the $\beta\beta$ calculations, whereas in the
second method one necessarily should check the consistency of the calculations
against the available beta-decay observables. This is a necessary procedure, not
warranting either a plus or a minus mark. In point five the similar behaviour of the
two discussed fitting methods comes, on one hand, from the fact that in both 
methods one fixes first
the pairing parameters by semiempirical pairing gaps. This is an essential step and
produces, for each single-particle space, the consistent quasiparticle mean field.
On the other hand, as the next step, both methods use experimental data to fit 
the $g_{\rm pp}$ parameter. This two-step fitting procedure is enough to eliminate
almost completely the dependence of the computed $0\nu\beta\beta$-decay rates on the 
size of the model space. 

The point number six is a very important one considering the actual computation
of the $0\nu\beta\beta$-decay rates. In the first procedure a separate 
$g_{\rm pp}$ analysis of higher multipoles can be performed, e.g., in the 
pf shell where data on beta decays of $2^-$ states are available. In the second 
method the same value of the $g_{\rm pp}$ parameter has to be assumed for all 
multipoles. Finally, it is to be noted that in the previous analysis
the axial-vector coupling constant, $g_{\rm A}$, has been assumed to be
roughly the same for both the $\beta$ and $\beta\beta$ decays. Since no
exhaustive studies of this matter have been performed, we take this assumption at
face value in this work. 

Concluding, the last line of Table~\ref{tab:table3} sums up the positive and negative 
points of each method. This final balance clearly supports the argument that 
the beta-decay fitting should be favoured, rather than the $2\nu\beta\beta$-decay 
fitting.

\end{document}